\documentclass[12pt,a4paper]{article}
\usepackage{amssymb,amsmath}
\usepackage{amsfonts}
\usepackage{enumerate}
\usepackage{a4}
\newtheorem{defi}{Definition}
\newtheorem{prop}{Proposition}
\newtheorem{theorem}{Theorem}
\newtheorem{lemma}{Lemma}
\newtheorem{cor}{Corollary}

\newtheorem{eg}{Example}
\begin{document}
 \baselineskip16pt
	
	\title{\bf Multi-dimensional Constacyclic Codes of Arbitrary Length  over Finite Fields }
	\author{Swati Bhardwaj ~ and ~ Madhu Raka
		\\ \small{\em Centre for Advanced Study in Mathematics}\\
		\small{\em Panjab University, Chandigarh-160014, INDIA}\\ \small emails: swatibhardwaj2296@gmail.com, mraka@pu.ac.in\\
		\date{}}
	\maketitle
\maketitle
\begin{abstract} Multi-dimensional cyclic code is a natural generalization of cyclic code. In an earlier paper we explored two-dimensional constacyclic codes over finite fields. Following the same technique, here we  characterize the algebraic structure of multi-dimensional constacyclic codes, in particular three-dimensional $(\alpha,\beta,\gamma)$- constacyclic codes of arbitrary length $s\ell k$ and their duals over a finite field  $\mathbb{F}_q$, where $\alpha,\beta,\gamma$ are non zero elements of $\mathbb{F}_q$. We give necessary and sufficient conditions for a three-dimensional $(\alpha,\beta,\gamma)$- constacyclic code to be self-dual. \vspace{2mm}
		\end{abstract}
	
\section{Introduction}
A multi-dimensional cyclic code or more precisely an $n$- D cyclic code of length $s_1s_2\cdots s_n$ over $\mathbb{F}_q$ is an ideal in the polynomial ring $\mathbb{F}_q[x_1,x_2,\cdots, x_n]/ \langle x_1^{s_1}-1, x_2^{s_2}-1, \cdots, x_n^{s_n}-1 \rangle $.  Because of their rich mathematical structure, involving Algebraic Geometry or ideals in a polynomial quotient ring, multi-dimensional cyclic codes are of great importance. Two-dimensional (2-D) cyclic codes  are used in daily technical applications
such as video encoding. The characterization for 2-D cyclic codes, for the first time was presented by  Ikai et al. \cite{Ikai} in 1975. Since the method was pure, it did not help decode these codes. After that  Imai \cite{Imai} introduced basic theories for binary 2-D cyclic codes using the concept of `common zero'. This concept  got  attention from  authors who were interested in decoding, see for example \cite{Sakata,Sakata1}.\vspace{2mm}

One of the main concerns about $n$- D cyclic codes is to find the related generator polynomials, because this enables us to investigate the structure of $n$- D cyclic codes and their duals. This procedure  helps to decode $n$- D cyclic codes also.\vspace{2mm}

In 2016,  Sepasdar and  Khashyarmanesh \cite{Sepas2016} obtained generator polynomials of
two-dimensional cyclic codes of length $s.2^k$ $(s_1=s, s_2=2^k)$ over $\mathbb{F}_{p^m} $ iteratively,  where $p$ is an odd prime. The authors state in the concluding remarks of their paper  \cite{Sepas2016} that their method does not work for arbitrary 2-D cyclic codes, not even when $s_1=3, s_2=3$.  In 2017, Sepasdar  \cite{Sepas2017}(unpublished) gave a  method for obtaining generator matrix of 2-D cyclic codes of arbitrary length $s_1s_2$, but this construction does not help much in yielding  numerical examples. \vspace{2mm}

Constacyclic codes over finite fields have a very significant role in the theory of error-correcting codes. A lot of work on constacyclic codes has been done in recent years, see for example \cite{BR, Chen, Dinh2, GR5, R, RKG}. Given nonzero elements $\alpha$ and $\beta$  of  $\mathbb{F}_q $, a two-dimensional $(\alpha,\beta )$-constacyclic code of length $s\ell$ is an ideal of the polynomial ring $\mathbb{F}_q[x, y]/ \langle x^s-\alpha, y^{\ell}-\beta \rangle $.  In 2018, Rajabi and Khashyarmanesh \cite {RajKhas}  investigated some repeated-root two-dimensional $(\alpha,\beta )$-constacyclic codes of length $2p^s.2^k$ over $\mathbb{F}_{p^m} $, where $p$ is an odd prime, using the structure of 2-D cyclic codes given in \cite{Sepas2016}.
\vspace{2mm}

 The authors   \cite{BR1} gave a novel method to characterize the algebraic structure of  2-D $(\alpha,\beta )$-constacyclic codes of arbitrary length $s\ell$ and their duals
over a finite field $\mathbb{F}_q $, using central primitive idempotents of the ring $\mathbb{F}_q[ y]/ \langle  y^{\ell}-\beta \rangle $. This method is quite different from that of \cite{RajKhas, Sepas2016, Sepas2017} and it does not require $s$ to be a multiple of $p$, the characteristic of $\mathbb{F}_q $ as in \cite {RajKhas}.   For $\alpha,\beta \in \{1,-1\}$, the necessary and sufficient conditions for a 2-D $(\alpha,\beta )$-constacyclic code to be self-dual are also given in \cite{BR1}.  As a consequence, a number of examples of  self-dual or isodual,  MDS or near MDS and quasi-twisted codes were given.\vspace{2mm}


There are much less results about general $n$-D cyclic codes. G$\ddot{\rm u}$neri et al. \cite{Gun2008} obtained a trace representation for multidimensional cyclic codes via Delsarte's theorem. This relates the weights of the codewords to the number of affine rational points of Artin Schreier type hypersurfaces over finite fields. In 2018, Lalasoa et al \cite{Lala1} generalized the method of Sepasdar \cite{Sepas2017} and constructed a basis of a 3-D
 cyclic code. Andriamifidisoa et al. \cite{And} and Lalasoa et al. \cite{Lala1}, \cite{Lala2} applied the method of Sepasder \cite{Sepas2017} to construct a basis of a 3-D cyclic code of arbitrary length and then they generalized this construction to a general $n$-D cyclic code. As is for 2-D codes, this construction does not help much in yielding numerical examples. \vspace{2mm}

  In this paper, following the method of \cite{BR1}, we study multi-dimensional constacyclic codes and their duals. In fact, the results of this paper are generalizations of the results of \cite{BR1} for 2-D $(\alpha,\beta )$-constacyclic codes of arbitrary length $s\ell$. Due to the ease in
visualization of the idea, we give the construction of generator polynomials of 3-D constacyclic codes and their duals. Generator matrices of a general $n$-D constacyclic code can be computed in a similar way. A 3-D $(\alpha,\beta,\gamma)$- constacyclic code  $\mathcal{C}$ of length $s\ell k$ is an ideal of the ring $\mathbb{F}_q[x, y,z]/ \langle x^s-\alpha, y^{\ell}-\beta, z^k-\gamma \rangle $. In Section 2.1, we study primitive central idempotents of the rings $\mathbb{F}_q[z]/\langle z^k - \gamma \rangle $ and  $\mathbb{F}_q[y]/\langle y^{\ell} - \beta \rangle $ and discuss some of their properties. Generator polynomials and generator matrices of 3-D $(\alpha,\beta,\gamma)$- constacyclic code $\mathcal{C}$ and its dual are obtained in Sections 2.2 and 2.3 respectively. In Section 2.4, we give  necessary and sufficient conditions for a 3-D constacyclic code to be self-dual for $\alpha, \beta, \gamma \in \{1,-1\}$.\vspace{2mm}

\section{Three Dimensional $(\alpha,\beta,\gamma)$ - constacyclic Codes of  length $s\ell k$}
 Firstly we recall the definition of a $\lambda$- quasi-twisted code over $\mathbb{F}_q$.
\begin{defi}\normalfont Let \vspace{2mm}

$\begin{array}{ll}a&=\big(a_{0,0},a_{1,0},\cdots,a_{m-1,0}|a_{0,1},\cdots,a_{m-1, 1}|\cdots|a_{0,n-1},\cdots,a_{m-1,n-1}\big)\vspace{2mm}\\& = \big(a^{(0)}|a^{(1)}|\cdots |a^{(n-1)}\big)\end{array}$ \vspace{1mm}

\noindent where $a^{(i)}= (a_{0,i},a_{1,i},\cdots,a_{m-1,i})$,   be a vector in  $\mathbb{F}_q^{mn}$ divided into $n$ equal parts each of length $m$. A linear code $C$ of length $mn$ over $\mathbb{F}_q$ is called a $\lambda$- quasi-twisted code of index $n$ if   $ \tau_{\lambda}(a)=\big(\lambda a^{(n-1)}|a^{(0)}|\cdots |a^{(n-2)}\big)\in C$ whenever $a\in C$.  When $\lambda=1$, $C$ is called quasi-cyclic code of index $n$. \end{defi}
\noindent Let $\mathcal{C}$ be a three dimensional $(\alpha,\beta,\gamma)$- constacyclic code i.e. $\mathcal{C}$ is an ideal of the ring $\mathcal{R}=\mathbb{F}_q[x, y,z]/ \langle x^s-\alpha, y^{\ell}-\beta, z^k-\gamma \rangle $. Each codeword $c$ in $\mathcal{C}$ has a unique polynomial representation
\begin{equation}\label{eq0}c(x,y,z)=\displaystyle \sum_{i=0}^{s-1} \sum_{j=0}^{\ell-1}\sum_{t=0}^{k-1}c_{i,j,t}~x^iy^jz^t.\end{equation}

\noindent For fixed $j,t, 0\leq j \leq \ell -1, 0\leq t \leq k-1, $ let $$c_x^{(y^j, z^t)} = (c_{0,j,t}, c_{1,j,t}, \cdots c_{s-1,j,t})$$
denote the $s$-tuple whose co-ordinates are coefficients of $x^i y^j z^t$, $0 \leq i \leq s-1$ in $c(x,y,z)$ given by (\ref{eq0}).
 The subscript $x$ in $c_x^{(y^j, z^t)}$ denotes that in this tuple the power $i$ in  $x^i$ is varying. Define,

$$c^{(z^t)} = \big(c_x^{(y^0, z^t)}~\vline~ c_x^{(y^1, z^t)}~\vline~ \cdots ~\vline~ c_x^{(y^{\ell -1}, z^t)}\big)$$

\noindent then, $c^{(z^t)}$ is a $s\ell$- tuple whose co-ordinates are all the coefficients of $z^t$ in $c(x,y,z).$

\noindent Similarly, let
$$c_y^{(x^i, z^t)} = (c_{i,0,t}, c_{i,1,t}, \cdots c_{i,\ell -1,t})$$
denote the $\ell$-tuple whose co-ordinates are coefficients of $x^i y^j z^t$, for fixed $i,t$ and  $j$ in $y^j$ is varying from $0$ to $ \ell-1$. Let

$$c^{(x^i)} = \big(c_y^{(x^i, z^0)}~\vline~ c_y^{(x^i, z^1)}~\vline~ \cdots ~\vline~ c_y^{(x^i, z^{k-1})}\big).$$

\noindent Then, $c^{(x^i)}$ is a $\ell k$- tuple whose co-ordinates are all the coefficients of $x^i$ in $c(x,y,z).$

\noindent In the same way, let
$$c_z^{(x^i, y^j)} = (c_{i,j,0}, c_{i,j,1}, \cdots c_{i,j,k-1})$$
denote the $k$-tuple whose co-ordinates are coefficients of $x^i y^j z^t$, for fixed $i,j$ and $t $ in $z^t$ varies from $0$ to  $ k-1$ . Define

$$c^{(y^j)} = \big(c_z^{(x^0, y^j)}~\vline~ c_z^{(x^1, y^j)}~\vline~ \cdots ~\vline~ c_z^{(x^{s-1}, y^j)}\big)$$

\noindent then, $c^{(y^j)}$ is a $s k$- tuple whose co-ordinates are all the coefficients of $y^j$ in $c(x,y,z).$\vspace{2mm}

\noindent For a 3-D constacyclic code  $\mathcal{C}$, let
\begin{equation}\label{Eq1}\mathcal{C}_1= \left\{  \big(c^{(x^0)}|c^{(x^1)}|\cdots|c^{(x^{s-1})}\big) :c(x,y,z) \in \mathcal{C} \right \}, \end{equation}
 \begin{equation}\label{Eq2}\mathcal{C}_2= \left\{  \big(c^{(y^0)}|c^{(y^1)}|\cdots|c^{(y^{\ell-1})}\big) :c(x,y,z) \in \mathcal{C} \right \}, \end{equation}
  \begin{equation}\label{Eq3}\mathcal{C}_3= \left\{  \big(c^{(z^0)}|c^{(z^1)}|\cdots|c^{(z^{k-1})}\big) :c(x,y,z) \in \mathcal{C} \right \}. \end{equation}
Clearly $\mathcal{C}_1$, $\mathcal{C}_2$ and $\mathcal{C}_3$ are linear codes in $\mathbb{F}_q^{s\ell k}$ and are permutation equivalent. \vspace{2mm}

 \noindent Note that $xc(x,y,z)$ corresponds to the codeword $ \big(\alpha c^{(x^{s-1})}| c^{(x^0)}|c^{(x^1)}|\cdots|c^{(x^{s-2})}\big)$,\\ $yc(x,y,z)$ corresponds to the codeword $\big(\beta c^{(y^{\ell-1})}|c^{(y^0)}|c^{(y^1)}|\cdots|c^{(y^{\ell-2})}\big)$ and $zc(x,y,z)$ corresponds to the codeword $ \big(\gamma c^{(z^{k-1})}|c^{(z^0)}|c^{(z^1)}|\cdots|c^{(z^{k-2})}\big)$.\vspace{2mm}

\noindent  By definitions of a 3-D code and of a $\lambda$- quasi-twisted code we immediately have

\begin{prop} $\mathcal{C}$ is a 3-D $(\alpha,\beta,\gamma )$-constacyclic code if and only if $\mathcal{C}_1$ is an $\alpha$- quasi-twisted code of index $s$, $\mathcal{C}_2$ is  a $\beta$- quasi-twisted code of index $\ell$ and $\mathcal{C}_3$ is  a $\gamma$- quasi-twisted code of index $k$.\end{prop}

 \noindent For a linear code $\mathcal{C}$ of length $n$ over $\mathbb{F}_q$, the dual code  $\mathcal{C}^\bot$ is defined as $\mathcal{C}^\bot =\{x\in \mathbb{F}_q^n~ |~ x \cdot y=0 ~ {\rm for ~ all~} y \in \mathcal{C}\}$, where $x\cdot y$ denotes the usual Euclidean inner product. It is well known that if $\mathcal{C}$ is $\lambda$ - constacyclic code over $\mathbb{F}_q$, then $\mathcal{C}^\bot$ is a $\lambda^{-1}$ - constacyclic code over $\mathbb{F}_q$. Rajabi and Khashyarmanesh \cite {RajKhas} showed that the dual of a two-dimensional $(\alpha,\beta )$-constacyclic code is a two-dimensional $(\alpha^{-1},\beta^{-1} )$-constacyclic code. A similar result holds for a  3-dimensional $(\alpha,\beta,\gamma )$-constacyclic code.

\begin{prop} The dual of a three-dimensional $(\alpha,\beta,\gamma)$-constacyclic code is a three-dimensional $(\alpha^{-1},\beta^{-1},\gamma^{-1} )$-constacyclic code.\end{prop}

\noindent {\bf Proof :} Let $\mathcal{C}_1, \mathcal{C}_2, \mathcal{C}_3$ be the corresponding linear codes in $\mathbb{F}_q^{s\ell k}$ as defined in (\ref{Eq1}), (\ref{Eq2}), and (\ref{Eq3}) respectively.
Let $\mathcal{C}_1^\perp, \mathcal{C}_2^\perp, \mathcal{C}_3^\perp$ be the duals of $\mathcal{C}_1, \mathcal{C}_2, \mathcal{C}_3$ in $\mathbb{F}_q^{s\ell k}$. \vspace{2mm}

\noindent To prove that $\mathcal{C}^\perp$ is $3$-dimensional $(\alpha^{-1}, \beta^{-1}, \gamma^{-1})$- constacyclic code, it is enough to prove that $\mathcal{C}_1^\perp$ is $\alpha^{-1}$- quasi twisted code , $\mathcal{C}_2^\perp$ is $\beta^{-1}$- quasi twisted code and $\mathcal{C}_3^\perp$ is $\gamma^{-1}$- quasi twisted code.

\noindent Let $a=\big( a^{(x^0)} ~\vline~ a^{(x^1)} ~\vline~ \cdots ~\vline~ a^{(x^{s-1})}   \big) \in \mathcal{C}_1^\perp , $ where $a(x,y,z) \in \mathcal{C}^\perp$
and $b=\big( b^{(x^0)} ~\vline~ b^{(x^1)} ~\vline~ \cdots ~\vline~ b^{(x^{s-1})}   \big)$ be an arbitrary element of $\mathcal{C}_1 $. Then $$\tau_\alpha^{s-1}(b) = \big(\alpha b^{(x^1)} ~\vline~ \alpha b^{(x^2)} ~\vline~ \cdots ~\vline~ \alpha b^{(x^{s-1})} ~\vline~ b^{(x^0)}   \big).$$

\noindent As $\mathcal{C}_1$ is $\alpha$- quasi twisted code , $\tau_\alpha^{s-1}(b) \in \mathcal{C}_1.$ Therefore, by definition of duality $a \cdot \tau_\alpha^{s-1}(b)=0.$ This gives
\begin{equation*}\begin{array}{ll}
&a^{(x^0)}\cdot\alpha b^{(x^1)} + a^{(x^1)}\cdot\alpha b^{(x^2)} + \cdots + a^{(x^{s-2})}\cdot\alpha b^{(x^{s-1})} + a^{(x^{s-1})}\cdot b^{(x^0)} = 0\vspace{2mm}\\
\Rightarrow &\alpha^{-1} a^{(x^{s-1})} \cdot b^{(x^0)} + a^{(x^0)} \cdot b^{(x^1)} + \cdots + a^{(x^{s-2})} \cdot b^{(x^{s-1})} = 0\vspace{2mm}\\
{\rm i.~e.} &\big( \alpha^{-1}a^{(x^{s-1})} ~\vline~ a^{(x^0)} ~\vline~ \cdots ~\vline~ a^{(x^{s-2})} \big) \cdot \big( b^{(x^0)} ~\vline~ b^{(x^1)} ~\vline~ \cdots ~\vline~  b^{(x^{s-1})} \big) =0 \vspace{2mm}\\
{\rm i.~e.} &\tau_{\alpha^{-1}}(a)\cdot b =0 ~{ \rm  ~for~ all~} b \in \mathcal{C}_1.
\end{array}
\end{equation*}

\noindent This gives $\tau_{\alpha^{-1}}(a) \in \mathcal{C}_1^\perp$ whenever $a \in \mathcal{C}_1^\perp$. Therefore, $\mathcal{C}_1^\perp$ is $\alpha^{-1}$- quasi twisted code.
\noindent Similarly, one gets that $\mathcal{C}_2^\perp$ is $\beta^{-1}$- quasi twisted code and $\mathcal{C}_3^\perp$ is $\gamma^{-1}$- quasi twisted code. \hfill $\square$\vspace{2mm}

\begin{defi}\normalfont Let
	\begin{equation*}
	\begin{array}{lll}
	c &=& \big( c^{(z^0)}|c^{(z^1)}| \cdots |c^{(z^{k-1})} \big)\vspace{2mm} \\
	&=& \big(c_x^{(y^0z^0)} ~\vline~ c_x^{(y^1z^0)} ~\vline~ \cdots ~\vline~ c_x^{(y^{\ell -1}z^0)} ~\vline~\cdots~ \cdots ~\vline~  c_x^{(y^0z^{k-1})} ~\vline~ c_x^{(y^1z^{k-1})} ~\vline~ \cdots ~\vline~ c_x^{(y^{\ell -1}z^{k-1})}     \big) \vspace{2mm}\\
	&=& \big(c_{0,0,0}, c_{1,0,0}, \cdots, c_{s-1,0,0} ~\vline~ c_{0,1,0}, c_{1,1,0}, \cdots, c_{s-1,1,0} ~\vline~ \cdots ~\vline~ c_{0,\ell-1,0}, c_{1,\ell-1,0}, \cdots, c_{s-1,\ell-1,0}~\vline~\vspace{1mm}\\
	&&\cdots~ \cdots ~\vline~  c_{0,0,k-1}, c_{1,0,k-1}, \cdots, c_{s-1,0,k-1} ~\vline~ c_{0,1,k-1}, c_{1,1,k-1}, \cdots, c_{s-1,1,k-1} ~\vline~\vspace{1mm}\\
	&& \cdots ~\vline~ c_{0,\ell-1,k-1}, c_{1,\ell-1,k-1}, \cdots, c_{s-1,\ell-1,k-1} \big)\in \mathcal{C}.
	\end{array}
	\end{equation*}
	
	\noindent  For $\alpha, \beta, \gamma \in \mathbb{F}_q^*$, define
	\begin{equation*}\begin{array}{lll}
	\tau^{1,0,0}_{\alpha, \beta, \gamma}(c)&=& \big( \alpha c_{s-1,0,0}, c_{0,0,0},  \cdots ,c_{s-2,0,0} ~\vline~\cdots ~\vline~  \alpha c_{s-1,\ell -1,0}, c_{0,\ell -1,0}, \cdots ,c_{s-2,\ell -1,0}~\vline~\vspace{1mm}\\
	&& \cdots ~\cdots ~\vline~   \alpha c_{s-1,0,k-1}, c_{0,0,k-1},  \cdots ,c_{s-2,0,k-1} ~\vline~\cdots ~ \vspace{1mm}\\
	&& ~\vline~\alpha c_{s-1,\ell -1,k-1}, c_{0,\ell -1,k-1}, \cdots ,c_{s-2,\ell -1,k-1} \big)\vspace{2mm}\\

	\tau^{0,1,0}_{\alpha, \beta, \gamma}(c)&=& \big( \beta c_{0,\ell-1,0}, \beta c_{1,\ell-1,0},  \cdots ,\beta c_{s-1,\ell-1,0} ~\vline~c_{0,0,0},c_{1,0,0},\cdots, c_{s-1,0,0}~\vline~\vspace{1mm}\\
	&&
	\cdots ~\vline~  c_{0,\ell -2,0}, c_{1,\ell -2,0}, \cdots ,c_{s-1,\ell -2,0}
	 ~\vline~\cdots ~\cdots ~\vline~
	 \beta c_{0,\ell-1,k-1}, \beta c_{1,\ell-1,k-1}, \vspace{1mm}\\ &&\cdots ,\beta c_{s-1,\ell-1,k-1} ~\vline~c_{0,0,k-1},c_{1,0,k-1},\cdots, c_{s-1,0,k-1}~\vline~
	 \cdots ~\vline~  c_{0,\ell -2,k-1},\vspace{1mm}\\
	 && c_{1,\ell -2,k-1}, \cdots ,c_{s-1,\ell -2,k-1} \big)\vspace{2mm}\\
&= &\big( \beta c_x^{(y^{\ell-1}z^0)}~\vline ~c_x^{(y^0z^0)}~\vline\cdots\vline ~c_x^{(y^{\ell-2}z^0)}~\vline \cdots \vline ~\beta c_x^{(y^{\ell-1}z^{k-1})}~\vline ~c_x^{(y^0z^{k-1})}~\vline \cdots \vline c_x^{(y^{\ell-2}z^{k-1})}\big)\vspace{2mm}\\	
	
	\tau^{0,0,1}_{\alpha, \beta, \gamma}(c)&=& \big(\gamma c_{0,0,k-1}, \gamma c_{1,0,k-1}, \cdots ,\gamma c_{s-1,0,k-1} ~\vline~ \gamma c_{0,1,k-1}, \gamma c_{1,1,k-1}, \cdots, \gamma c_{s-1,1,k-1} ~\vline~\vspace{1mm}\\
	&& \cdots ~\vline~ \gamma c_{0,\ell-1,k-1}, \gamma c_{1,\ell-1,k-1}, \cdots, \gamma c_{s-1,\ell-1,k-1}     ~\vline~  c_{0,0,0}, c_{1,0,0}, \cdots ,c_{s-1,0,0} ~\vline~\vspace{1mm}\\
	&&  \cdots ~\vline~ c_{0,\ell-1,0}, c_{1,\ell-1,0}, \cdots c_{s-1,\ell-1,0}~ \vline~ \cdots ~\vline  c_{0,\ell-1,k-2}, c_{1,\ell-1,k-2}, \cdots,  c_{s-1,\ell-1,k-2}     ~\vline~  \big)\vspace{2mm}\\&=&
\big( \gamma c_x^{(y^{0}z^{k-1})}~\vline ~\gamma c_x^{(y^1z^{k-1})}~\vline\cdots\vline ~\gamma c_x^{(y^{\ell-1}z^{k-1})}~\vline ~ c_x^{(y^{0}z^{0})}~\vline \cdots \vline ~c_x^{(y^{\ell-1}z^{0})}~\vline \cdots \vline c_x^{(y^{\ell-1}z^{k-2})}\big)\vspace{2mm}\\&=&\big(\gamma c^{(z^{k-1})}| c^{(z^{0})}|
\cdots| c^{(z^{k-2})}\big).
	\end{array}
	\end{equation*}
\end{defi}\vspace{2mm}

\begin{prop}
	Let $f(x,y,z) , g(x,y,z) \in \mathbb{F}_q[x,y,z]/ \langle x^s-\alpha, y^\ell-\beta, z^k-\gamma \rangle$ and let codewords corresponding to $f(x,y,z)$ and $g(x,y,z)$ be
	\begin{equation*}\begin{array}{ll}
	\underline{a}&= \big(a^{(z^0)} ~\vline~ a^{(z^1)} ~\vline~ \cdots ~\vline~ a^{(z^{k-1})}  \big)\\
	 &= \big(a_x^{(y^0z^0)} ~\vline~ a_x^{(y^1z^0)} ~\vline~ \cdots ~\vline~ a_x^{(y^{\ell -1}z^0)} ~\vline~\cdots \cdots ~\vline~  a_x^{(y^0z^{k-1})} ~\vline~ a_x^{(y^1z^{k-1})} ~\vline~ \cdots ~\vline~ a_x^{(y^{\ell -1}z^{k-1})}     \big)
	\end{array}
	\end{equation*}
	\noindent and,
	\begin{equation*}\begin{array}{ll}
	\underline{b}&= \big(b^{(z^0)} ~\vline~ b^{(z^1)} ~\vline~ \cdots ~\vline~ b^{(z^{k-1})}  \big)\\
	&= \big(b_x^{(y^0z^0)} ~\vline~ b_x^{(y^1z^0)} ~\vline~ \cdots ~\vline~ b_x^{(y^{\ell -1}z^0)} ~\vline~\cdots ~ \cdots ~\vline~  b_x^{(y^0z^{k-1})} ~\vline~ b_x^{(y^1z^{k-1})} ~\vline~ \cdots ~\vline~ b_x^{(y^{\ell -1}z^{k-1})}     \big)
	\end{array}
	\end{equation*}
	\noindent respectively.
	
	\noindent Then, $f(x,y,z)g(x,y,z)=0$ in $\mathbb{F}_q[x,y,z]/ \langle x^s-\alpha, y^\ell-\beta, z^k-\gamma \rangle$ if and only if
	$\underline{a}$ is orthogonal to $\underline{b}^*=\big(  {b_x^*}^{(y^{\ell -1}z^{k-1})}~\vline~ \cdots ~\vline~  {b_x^*}^{(y^0z^{k-1})} ~\vline~ \cdots ~\cdots ~\vline~ {b_x^*}^{(y^1z^0)} ~\vline~ {b_x^*}^{(y^0z^0)}      \big)$
	and all its $(\alpha^{-1},\beta^{-1},\gamma^{-1})$ - constacyclic shifts, where ${b_x^*}^{(y^j z^t)} = (b_{s-1,j,t},b_{s-2,j,t}, \cdots,b_{0,j,t} )$.
\end{prop}

\noindent \textbf{Proof:} For convenience, we give a proof for $s=\ell=k=2$. Also we skip commas in subscripts of  $c_{i,j,t}$. For, $c= (c^{(z^0)} ~\vline~ c^{(z^1)}) = (c_x^{(y^0z^0)}  ~\vline~ c_x^{(y^1z^0)}  ~\vline~ c_x^{(y^0z^1)}   ~\vline~ c_x^{(y^1z^1)}  )  =( c_{000},c_{100} ~\vline~  c_{010},c_{110} ~\vline~  c_{001},c_{101} ~\vline~ c_{011},c_{111}   ) $ we have
\begin{equation*}\begin{array}{ll}
\tau^{~001}_{\alpha^{-1}\beta^{-1}\gamma^{-1}}(c) &= \big( \gamma^{-1} c^{(z^1)} ~\vline~ c^{(z^0)}   \big)\\
&=\big( \gamma^{-1}c_{001}, \gamma^{-1}c_{101}~\vline~   \gamma^{-1}c_{011}, \gamma^{-1}c_{111}~\vline~  c_{000}, c_{100}~\vline~   c_{010}, c_{110}    \big) \vspace{4mm}\\

\tau^{~010}_{\alpha^{-1}\beta^{-1}\gamma^{-1}}(c) &= \big( \beta^{-1}c_{010}, \beta^{-1} c_{110}~\vline~c_{000}, c_{100}~\vline~\beta^{-1}c_{011}, \beta^{-1}c_{111}~\vline~c_{001}, c_{101}\big) \vspace{4mm}\\

\tau^{~100}_{\alpha^{-1}\beta^{-1}\gamma^{-1}}(c) &= \big( \alpha^{-1}c_{100},  c_{000}~\vline~  \alpha^{-1}c_{110}, c_{010}~\vline~\alpha^{-1}c_{101}, c_{001}~\vline~ \alpha^{-1}c_{111}, c_{011}\big)
\end{array}
\end{equation*}

\noindent Let $f(x,y,z) g(x,y,z) =h(x,y,z) = \sum h_{ijt} x^i y^j z^t $, $i.e.$ \\
 $$\begin{array}{ll}h(x,y,z) = &\big( a_{000} x^0 y^0 z^0 + a_{100} x^1 y^0 z^0 + a_{010} x^0 y^1 z^0 +  a_{110} x^1 y^1 z^0 + a_{001} x^0 y^0 z^1 \\& ~~~~~~~+ a_{101} x^1 y^0 z^1 + a_{011} x^0 y^1 z^1 + a_{111} x^1 y^1 z^1 \big) \\& \big( b_{000} x^0 y^0 z^0 + b_{100} x^1 y^0 z^0 + b_{010} x^0 y^1 z^0 +  b_{110} x^1 y^1 z^0 + b_{001} x^0 y^0 z^1\\&~~~~~~~ + b_{101} x^1 y^0 z^1 + b_{011} x^0 y^1 z^1 + b_{111} x^1 y^1 z^1      \big)\end{array}$$

\noindent Then for $k_1, k_2, k_3 \in \{0,1\} \pmod 2$, coefficient of $x^{k_1} y^{k_2} z^{k_3}$ is

\begin{equation*}
\begin{array}{ll}
h_{k_1k_2k_3} =&
\displaystyle \sum_{\substack {i+i'=k_1\\j+j'=k_2\\t+t'=k_3}}a_{ijt} b_{i'j't'} +
\displaystyle\sum_{\substack {i+i'=s+k_1\\j+j'=k_2\\t+t'=k_3}}\alpha a_{ijt}b_{i'j't'} +

\displaystyle\sum_{\substack {i+i'=k_1\\j+j'=\ell +k_2\\t+t'=k_3}}\beta a_{ijt} b_{i'j't'}+\vspace{4mm}\\
&\displaystyle\sum_{\substack {i+i'=k_1\\j+j'=k_2\\t+t'=k+k_3}}\gamma a_{ijt} b_{i'j't'}  +\displaystyle\sum_{\substack {i+i'=s+k_1\\j+j'=\ell +k_2\\t+t'=k_3}}\alpha \beta a_{ijt} b_{i'j't'} + \displaystyle\sum_{\substack {i+i'=s+k_1\\j+j'=k_2\\t+t'=k+k_3}}\alpha \gamma a_{ijt} b_{i'j't'} +\vspace{4mm}\\
& \displaystyle\sum_{\substack {i+i'=k_1\\j+j'=\ell +k_2\\t+t'=k+k_3}}\beta \gamma a_{ijt} b_{i'j't'} + \displaystyle\sum_{\substack {i+i'=s+k_1\\j+j'=\ell +k_2\\t+t'=k+k_3}}\alpha \beta \gamma a_{ijt} b_{i'j't'}
\end{array}
\end{equation*}

\noindent For $k_1, k_2, k_3 = 0,1 \pmod 2$, one finds  that
\begin{equation*}
\begin{array}{l}
h_{k_1k_2k_3} = \alpha^{k_1+1} \beta^{k_2+1} \gamma^{k_3+1}\underline{a}  \cdot ~\tau^{k_1+1, k_2+1, k_3+1}_{\alpha^{-1}\beta^{-1}\gamma^{-1}}\big(
b_{111},b_{011},b_{101},b_{001},b_{110},b_{010},b_{100},b_{000}\big)
\end{array}
\end{equation*}

\noindent For example \begin{equation*}\begin{array}{lll}h_{100}&=& {\rm coefficient ~ of ~~} x^1y^0z^0 {\rm ~ in ~} h(x,y,z)\\&=& a_{000}b_{100} +a_{100}b_{000} +\beta a_{010}b_{110} +\beta a_{110}b_{010} +\gamma a_{001}b_{101} +\gamma a_{101}b_{001}\vspace{1mm}\\&& +\beta \gamma  a_{011}b_{111} +\beta\gamma a_{111}b_{011}\vspace{2mm}\\&=&\big(a_{000}, a_{100}, a_{010}, a_{110}, a_{001}, a_{101}, a_{011}, a_{111} \big)\vspace{1mm}\\ && \cdot \big(b_{100}, b_{000}, \beta b_{110}, \beta b_{011}, \gamma b_{101}, \gamma b_{001}, \beta \gamma b_{111}, \beta \gamma b_{011}   \big) \vspace{2mm}\\&=&\beta \gamma~ \underline{a} \cdot \big(\beta^{-1}\gamma^{-1}b_{100}, \beta^{-1}\gamma^{-1}b_{000},\gamma^{-1}  b_{110}, \gamma^{-1} b_{011},\beta^{-1} b_{101}, \beta^{-1} b_{001},   b_{111},  b_{011}   \big) \vspace{2mm}\\&=& \beta \gamma~ \underline{a} \cdot~ \tau^{~0,1,1}_{\alpha^{-1}\beta^{-1}\gamma^{-1}}\big(b_{111},b_{011},b_{101},b_{001},b_{110},b_{010},b_{100},
b_{000}\big)\vspace{2mm}\\&=& \beta \gamma ~\underline{a}   \cdot~ \tau^{~0,1,1}_{\alpha^{-1}\beta^{-1}\gamma^{-1}}\big( b_x^{*(y^1z^1)}|b_x^{*(y^0z^1)}| b_x^{*(y^1z^0)}|b_x^{*(y^0z^0)}\big)\end{array}\end{equation*}

\begin{equation*}\begin{array}{lll}h_{000}&=& {\rm coefficient ~ of ~~} x^0y^0z^0 \\&=& a_{000}b_{000} +\alpha a_{100}b_{100} +\beta a_{010}b_{010} +\alpha \beta a_{110}b_{110} +\gamma a_{001}b_{001} +\alpha \gamma a_{101}b_{101}\vspace{1mm}\\&& +\beta \gamma  a_{011}b_{011} +\alpha \beta\gamma a_{111}b_{111}\vspace{2mm}\\&=&\big(a_{000}, a_{100}, a_{010}, a_{110}, a_{001}, a_{101}, a_{011}, a_{111} \big)\vspace{1mm}\\ && \cdot \big(b_{000},\alpha b_{100}, \beta b_{010},\alpha \beta b_{110}, \gamma b_{001}, \alpha \gamma b_{101}, \beta \gamma b_{011}, \alpha \beta \gamma b_{111}   \big) \vspace{2mm}\\&=& \alpha \beta \gamma ~\underline{a}\cdot \big(\alpha ^{-1}\beta^{-1}\gamma^{-1}b_{000}, \beta^{-1}\gamma^{-1}b_{100},\alpha ^{-1}\gamma^{-1}  b_{010}, \gamma^{-1} b_{110},\alpha ^{-1}\beta^{-1} b_{001}, \\ &&~~~~~~~~~~~~~~~~~~~~~\beta^{-1} b_{101},  \alpha ^{-1} b_{011},  b_{111}   \big) \vspace{2mm}\\&=&\alpha \beta \gamma ~\underline{a}  \cdot~ \tau^{~1,1,1}_{\alpha^{-1}\beta^{-1}\gamma^{-1}}\big(b_{111},b_{011},b_{101},b_{001},b_{110},b_{010},b_{100},b_{000}
\big)\vspace{2mm}\\&=&\alpha \beta \gamma ~\underline{a}   \cdot~ \tau^{~1,1,1}_{\alpha^{-1}\beta^{-1}\gamma^{-1}}\big(\underline{b}^*\big)\end{array}\end{equation*}

\noindent Similarly one can check others as well. Therefore, $h_{k_1k_2k_3}=0$ if and only if
 $$\underline{a}   \cdot~ \tau^{~k_1+1,k_2+1,k_3+1}_{\alpha^{-1}\beta^{-1}\gamma^{-1}}\big( \underline{b}^*\big)=0,$$

\noindent  if and only if $\underline{a}$ ~is orthogonal to $\underline{b}^*$
and all its $(\alpha^{-1},\beta^{-1},\gamma^{-1})$- constacyclic shifts. \hfill $\square$ \vspace{3mm}

 Let $S$ be a non-empty subset of a commutative ring $R$. The annihilator of $S$, denoted by ann$(S)$, is the set ann$(S)= \{f \in R  : fg=0 {\rm ~for~ all~} g \in S\}$. Then ann$(S)$ is an ideal of $R$. For a polynomial $f(x)$ with $\deg(f(x))=k$, its reciprocal is defined as $f^\ast(x)=x^kf(1/x)$. The reciprocal of the zero polynomial is the zero polynomial itself. For any set S of polynomials over $\mathbb{F}_q$, we use the notation $S^\ast=\{ f^\ast : f\in S\}$. If $C$ is a $\lambda$-constacyclic code of length $n$ over $\mathbb{F}_q$ generated by $g(x)$, then
$C^\perp$, which is $\lambda^{-1}$-constacyclic code, is generated by 	$h^*(x)$ where $x^n-\lambda= g(x)h(x)$.
\begin{prop}
	Suppose that $\alpha,\beta, \gamma \in \{1,-1\}$. Let $\mathcal{C}$ be a three-dimensional $(\alpha,\beta, \gamma )$-constacyclic code, then $\mathcal{C}^\perp$ is also a three-dimensional $(\alpha,\beta, \gamma )$-constacyclic code and $\mathcal{C}^\perp = ({\rm ann}(\mathcal{C}))^\ast$, also denoted as   ann$^\ast(\mathcal{C})$.
	
\end{prop}
\noindent \textbf{Proof:} Proof follows from the above two propositions. \vspace{2mm}\\

\subsection{Primitive central idempotents and their properties}
In this section  we study primitive central idempotents of the rings $\mathbb{F}_q[z]/\langle z^k - \gamma \rangle $ and  $\mathbb{F}_q[y]/\langle y^{\ell} - \beta \rangle $ and discuss some of their properties.\vspace{2mm}

Let  $\mathcal{R}$ denote the polynomial ring $\mathbb{F}_q[x,y,z]/\langle x^s-\alpha,y^{\ell}-\beta,z^k-\gamma \rangle$.
Let $r$ be the order of $\gamma$ in $\mathbb{F}_q$ so that $\gamma^r=1$ and $r | (q-1)$.
 Let $\omega$ be a ${rk}^{th}$ root of unity such that $\omega^k=\gamma$. Assume that $q \equiv 1 \pmod {rk}$  so that $\omega \in \mathbb{F}_q$, since for some integer $k'$, $\omega^{q-1}=\omega^{r k k'}=\gamma^{rk'}=1$.  Then,
\begin{equation*}
z^{rk}-1=(z-1)(z-\omega)(z-\omega^2) \cdots (z-\omega^{rk-1}).
\end{equation*}
\noindent Define
\begin{equation}\label{eq5}\begin{array}{ll}
\xi_0(z) &=\displaystyle \frac{(z-\omega)(z-\omega^2)(z-\omega^3)\cdots(z-\omega^{rk-1})}{(1-\omega)(1-\omega^2)(1-\omega^3)\cdots
(1-\omega^{rk-1})}, \vspace{2mm}\\
\xi_1(z) &=\displaystyle \frac{(z-1)(z-\omega^2)(z-\omega^3)\cdots(z-\omega^{rk-1})}{(\omega-1)(\omega-\omega^2)(\omega-\omega^3)
\cdots(\omega-\omega^{rk-1})}, \vspace{2mm}\\
&\vdots \vspace{2mm}\\
\xi_t(z) &=\displaystyle \frac{(z-1)(z-\omega)\cdots(z-\omega^{t-1})(z-\omega^{t+1})\cdots(z-\omega^{rk-1})}{(\omega^t-1)
(\omega^t-\omega)\cdots(\omega^t-\omega^{t-1})(\omega^t-\omega^{t+1})\cdots(\omega^t-\omega^{rk-1})},\vspace{2mm}\\
&\vdots \vspace{2mm}\\
\xi_{rk-1}(z) &=\displaystyle \frac{(z-1)(z-\omega)(z-\omega^2)\cdots(z-\omega^{rk-2})}{(\omega^{rk-1}-1)(\omega^{rk-1}-\omega)
(\omega^{rk-1}-\omega^2)\cdots(\omega^{rk-1}-\omega^{rk-2})}.
\end{array}\end{equation}

\noindent Then  $\xi_0(z),\xi_1(z),\cdots,\xi_{rk-1}(z)$ are primitive central idempotents in $\mathbb{F}_q[z]/\langle z^{rk}-1\rangle $ $i.e.$ $\xi_0(z)+\xi_1(z)+ \cdots +\xi_{rk-1}(z)=1$ and $\xi_t(z) \xi_{t'}(z)= \delta_{t,t'} \xi_t(z)$ for $t,t' \in \{0,1,2, \cdots ,rk-1 \}$, where $\delta_{t,t'}$ is the Kronecker delta function. For a proof of it see \cite{GR2}. Also, $\xi_t(\omega^t)=1$ and $\xi_t(\omega^{t'})=0$ for $t \neq t'$.
The following Lemmas 1-4 are similar to \cite{BR1}, therefore proofs are skipped.

\begin{lemma} \label{lem1} \normalfont
	For $t=0,1,2,\cdots,rk-1$, we have $$\xi_t(z)=\displaystyle \frac{1}{rk} \Big( 1+\omega^{rk-t}z+(\omega^{rk-t}z)^2+\cdots+(\omega^{rk-t}z)^{rk-1} \Big).$$
\end{lemma}

\begin{lemma} \label{lem2} \normalfont
	$\xi_{1+tr}(z) z^{t'} = (\omega^{1+tr})^{t'} \xi_{1+tr}(z)$, for $~~t,t' \in \{ 0,1,2, \cdots ,k-1\} $.
\vspace{2mm}\end{lemma}

\noindent   Following the notations of \cite{GR5}, let $\mathbb{Z}_{k r}$ denote the residue ring $\mathbb{Z}$ modulo $k r$ and $\mathbb{Z}_{k r}^*$ denote the multiplicative group consisting of units of $\mathbb{Z}_{k r}$. Let
$$P_{k,\gamma} =1+r\mathbb{Z}_{k r}= \{1+rt : t=0,1,\cdots,k-1\} .$$
Then $\omega^i, i\in P_{k,\gamma}$ are all the roots of $z^k-\gamma$. For $a\in \mathbb{Z}_{k r}^*\cap P_{k,\gamma}$, a multiplier $\mu_a$ is a map  from  $P_{k,\gamma} \rightarrow P_{k,\gamma}$ defined as $\mu_a(i)\equiv ai ({\rm mod~} k r)$. It is  extended on $\mathbb{F}_q[z]/\langle z^k-\gamma\rangle$ by defining  $\mu_a(c(z))\equiv c(z^a) ({\rm mod~} z^k-\gamma)$.\vspace{2mm}\\
Since $\gcd(k,q)=1$ and $r|q-1$, it follows that $q\in \mathbb{Z}_{k r}^*\cap P_{k,\gamma}$. Clearly
  $P_{k,\gamma}$ is $\mu_q$-invariant and $P_{k,\gamma}$ is a union of $q$-cyclotomic cosets modulo $k r$. For any $q$-cyclotomic coset $Q$ with in $P_{k,\gamma}$,
$ M_Q(z)= \prod_{i\in Q}(z-\omega^i)$ is irreducible in $\mathbb{F}_q[z]$ and
\begin{equation}\begin{array}{ll}
z^k-\gamma&=\displaystyle\prod_Q M_Q(z)=\prod_{i\in P_{k,\gamma}}(z-\omega^i)\\&=(z-\omega)(z-\omega^{1+r})(z-\omega^{1+2r})\cdots(y-\omega^{1+(k-1)r}).
\vspace{2mm}\end{array}\end{equation}

\noindent Let $z^{rk}-1=(z^k-\gamma)Q(z)$, where $Q(z)= \displaystyle\prod_{i \notin P_{k,\gamma}} \left(z-\omega^i\right)$.
Define
\begin{equation}\label{eq7}\begin{array}{ll}
\zeta_0(z) &=\displaystyle \frac{(z-\omega^{1+r})(z-\omega^{1+2r})\cdots(z-\omega^{1+(k-1)r})}{(\omega-\omega^{1+r})(\omega-\omega^{1+2r})\cdots
	(\omega-\omega^{1+(k-1)r})}, \vspace{2mm}\\
\zeta_1(z) &=\displaystyle \frac{(z-\omega)(z-\omega^{1+2r})\cdots(z-\omega^{1+(k-1)r})}{(\omega^{1+r}-\omega)(\omega^{1+r}-\omega^{1+2r})\cdots
	(\omega^{1+r}-\omega^{1+(k-1)r})}, \vspace{2mm}\\
&\vdots \vspace{2mm}\\

\zeta_{k-1}(z) &=\displaystyle \frac{(z-\omega)(z-\omega^{1+r})\cdots(z-\omega^{1+(k-2)r})}{(\omega^{1+(k-1)r}-\omega)(\omega^{1+(k-1)r}-\omega^{1+r})\cdots
	(\omega^{1+(k-1)r}-\omega^{1+(k-2)r})}.
\end{array}\end{equation}

\noindent Then  $\zeta_0(z),\zeta_1(z),\cdots,\zeta_{k-1}(z)$ are primitive central idempotents in $\mathbb{F}_q[z]/\langle z^{k}-\gamma \rangle $.

\noindent Note that
\begin{equation} \label{eq8}{\rm If~} r=1, ~\zeta_{t}(z)=\xi_{t+1}(z) {\rm ~and ~}  \zeta_{k-1}(z)=\xi_0(z) \end{equation}

\begin{lemma} \label{lem3}\normalfont
	For $t=0,1,2,\cdots,k-1$ we have $\xi_{1+tr}(z)=\zeta_{t}(z) \left(  \frac{Q(z)}{c_{t}} \right) $ for some constant $c_{t}$ in $\mathbb{F}_q^*$.
\end{lemma}
Note that
\begin{equation}\label{eq9}\zeta_t(\omega^{1+tr})=1 {\rm ~and ~} \zeta_t(\omega^{1+t'r})=0 {\rm ~ for ~} t \neq t'.\end{equation}

\begin{lemma}\label{lem4} \normalfont
	$\zeta_{t}(z) z^{t'} = \big(\omega^{1+tr}\big)^{t'} \zeta_{t}(z)$ for $t,t' \in \{ 0,1,2, \cdots ,k-1\} $.
\end{lemma}

\begin{lemma} \label{lem5} \normalfont The reciprocal polynomials of $\zeta_{t}(z)$, for $t=0,1,\cdots,k-1,$ are given by
	\begin{equation}\label{eq10} \zeta_{t}^\ast(z) = \left\{ \begin{array}{lll}b_{t} ~\zeta_{k-2-t}(z) & {\rm if} & \gamma=1 \\
	b_{t}~ \zeta_{k-1-t}(z) & {\rm if} & \gamma=-1 \end{array}\right.\end{equation}
	for some constant $b_{t} \in \mathbb{F}_q^\ast$ with the understanding that for $\gamma=1$ and $t=k-1$, $\zeta_{k-2-t}(z)=\zeta_{-1}(z)=\xi_0(z)=\zeta_{k-1}(z)$ .
\end{lemma}
\noindent {\bf Proof :} When $\gamma=1$, we have $r=1$. Then by definition $\zeta_t(z)=\xi_{t+1}(z)$.
As $\xi_{t+1}(z)= \frac{1}{a_{t}}\frac{z^{k}-1}{z-\omega^{t+1}}$, for some constant $a_{t}$, and $\omega^{k}=1$, we have  $$\zeta_t^\ast(z)=\xi_{t+1}^\ast(z)=\frac{1}{a_{t}\omega^{t+1}}\frac{z^{k}-1}{(z-\omega^{-(t+1)})}=b_{t} ~\xi_{k-t-1}(z)= b_{t} ~\zeta_{k-t-2}(z)$$ for some constant $b_{t}$.\vspace{2mm}

When  $\gamma=-1$, we have $r=2$. As $\zeta_{t}(z)= \frac{1}{a_{t}}\cdot\frac{z^{k}+1}{z-\omega^{1+2t}}$, for some constant $a_{t}$, we get
$$\zeta_{t}^\ast(z)=\frac{1}{a_{t}~\omega^{1+2t}}\frac{z^{k}+1}{(z-\omega^{-(1+2t)})}=b_{t}~\frac{z^{k}+1}
{z-\omega^{1+2(k-1-t)}}=b_{t} ~\zeta_{k-1-t}(z)$$ for some constant $b_{t}$, since $\omega^{2k}=1$. \hfill $\Box$
\vspace{2mm}\\

Similarly, let $r'$ be the order of $\beta$ in $\mathbb{F}_q$ so that $\beta^{r'}=1$ and $r' | (q-1)$. Let $\theta$ be the $r'\ell ^ {th}$ root of unity such that $\theta^\ell = \beta$. Assume that   $q \equiv 1 \pmod {r'\ell}$  so that $\theta \in \mathbb{F}_q$.  We  define $\chi_j(y)$, the primitive central idempotents  in $\mathbb{F}_q[y]/\langle y^{r'\ell}-1\rangle $ similar to $\xi_t(z)$ as defined in (\ref{eq5}) and then define primitive central idempotents  $\eta_j(y)$ in $\mathbb{F}_q[y]/\langle y^\ell-\beta \rangle $  as in (\ref{eq7}). So we have
\begin{equation*}\begin{array}{ll}
\eta_0(y)&=\displaystyle \frac{(y-\theta^{1+r'})(y-\theta^{1+2r'})\cdots(y-\theta^{1+(\ell-1)r'})}{(\theta-\theta^{1+r'})(\theta-\theta^{1+2r'})
\cdots(\theta-\theta^{1+(\ell-1)r'})}, \vspace{2mm}\\
\eta_{1}(y)&=\displaystyle \frac{(y-\theta)(y-\theta^{1+2r'})\cdots(y-\theta^{1+(\ell-1)r'})}{(\theta^{1+r'}-\theta)(\theta^{1+r'}-\theta^{1+2r'})
\cdots(\theta^{1+r'}-\theta^{1+(\ell-1)r'})}, \vspace{2mm}\\
&\vdots \vspace{2mm}\\
\eta_{\ell-1}(y)&=\displaystyle \frac{(y-\theta)(y-\theta^{1+r'})\cdots(y-\theta^{1+(\ell-2)r'})}{(\theta^{1+(\ell-1)r'}-\theta)(\theta^{1+(\ell-1)r'}
-\theta^{1+r'})\cdots(\theta^{1+(\ell-1)r'}-\theta^{1+(\ell-2)r'})}.
\end{array}\end{equation*}

\noindent  We have similar results for $\chi_j(y)$ and $\eta_j(y)$ as obtained in Lemmas $\ref{lem1}, \ref{lem2}, \cdots, \ref{lem5}$. In particular
\begin{lemma} \label{lem6} \normalfont The reciprocal polynomials of $\eta_{j}(y)$, for $j=0,1,\cdots,\ell-1,$ are given by
	\begin{equation}\label{eq11} \eta_{j}^\ast(y) = \left\{ \begin{array}{lll}c_{j} ~\eta_{\ell-2-j}(y) & {\rm if} & \beta=1 \\
	c_{j}~ \eta_{\ell-1-j}(y) & {\rm if} & \beta=-1 \end{array}\right.\end{equation}
	for some constant $c_{j} \in \mathbb{F}_q^\ast$ with the understanding that for $\beta=1$ and $j=\ell-1$, $\eta_{\ell-2-j}(y)=\eta_{-1}(y)=\chi_0(y)=\eta_{\ell-1}(y)$.
\end{lemma}

\noindent Similar to (\ref{eq9}), we have

\begin{equation}\label{eq12}\eta_j(\theta^{1+jr'})=1 {\rm ~and~} \eta_j(\theta^{1+j'r'})=0 {\rm ~for ~} j \neq j'.\end{equation}

\subsection{Generator matrix}
In this section we  obtain generators of a three-dimensional $(\alpha,\beta,\gamma)$ - constacyclic code of arbitrary length $s\ell k$.

\noindent Let $\mathcal{C}$ be a three dimensional $(\alpha, \beta, \gamma)$ - constacyclic code i.e. $\mathcal{C}$ is an ideal of the ring $\mathcal{R}=\mathbb{F}_q[x,y,z]/\langle x^s-\alpha,y^\ell-\beta,z^k-\gamma \rangle$.
 Define,

\begin{equation*} \begin{array}{ll}
I_0 &= \{ f(x,y) \in \mathbb{F}_q[x,y]/\langle x^s-\alpha,y^\ell-\beta \rangle : \zeta_0(z)f(x,y) \in \mathcal{C} \} \vspace{2mm}\\
I_1 &= \{ f(x,y) \in \mathbb{F}_q[x,y]/\langle x^s-\alpha,y^\ell-\beta \rangle : \zeta_{1}(z)f(x,y) \in \mathcal{C} \} \vspace{2mm}\\
I_2 &= \{ f(x,y) \in \mathbb{F}_q[x,y]/\langle x^s-\alpha,y^\ell-\beta \rangle : \zeta_{2}(z)f(x,y) \in \mathcal{C} \} \vspace{2mm}\\
&\vdots \vspace{2mm}\\
I_{k-1} &= \{ f(x,y) \in \mathbb{F}_q[x,y]/\langle x^s-\alpha,y^\ell-\beta \rangle : \zeta_{k-1}(z)f(x,y) \in \mathcal{C} \}.
\end{array}\end{equation*}

\noindent Then $I_0,I_1,\cdots,I_{k-1}$ are ideals in the ring $\mathbb{F}_q[x,y]/\langle x^s-\alpha,y^\ell-\beta \rangle$, i.e.  $I_0,I_1,\cdots,I_{k-1}$ are two dimensional $(\alpha, \beta)$ - constacyclic codes.\\

\noindent For each $t$, $0\leq t \leq k-1$,  from Theorem 1 of \cite{BR1} , there exist polynomials $p_j^{(t)}(x)$, $0\leq j \leq \ell-1$, such that $p_j^{(t)}(x)|x^s-\alpha$ and the ideals $I_t$ are generated by  $$I_t=\langle \eta_0(y) p_0^{(t)}(x), \eta_1(y) p_1^{(t)}(x),\cdots,\eta_{\ell-1}(y) p_{\ell-1}^{(t)}(x) \rangle.$$
where $\eta_0(y),\eta_{1}(y),\cdots,\eta_{\ell-1}(y)$ are primitive central idempotents in $\mathbb{F}_q[y]/\langle y^\ell-\beta \rangle $.

\begin{theorem} \label{th1} Let $\mathcal{C}$ be an ideal in the ring $\mathcal{R}=\mathbb{F}_q[x,y,z]/\langle x^s-\alpha,y^\ell-\beta,z^k-\gamma \rangle$, then
\begin{equation}\label{eq13}
\begin{array}{ll}
\mathcal{C} = \big\langle &\zeta_0(z)\eta_0(y)p_0^{(0)}(x),  \cdots , \zeta_0(z)\eta_{\ell-1}(y)p_{\ell-1}^{(0)}(x),\\
&\zeta_1(z)\eta_0(y)p_0^{(1)}(x),  \cdots , \zeta_1(z)\eta_{\ell-1}(y)p_{\ell-1}^{(1)}(x),\\
& \vdots\\
& \zeta_{k-1}(z)\eta_0(y)p_0^{(k-1)}(x),  \cdots , \zeta_{k-1}(z)\eta_{\ell-1}(y)p_{\ell-1}^{(k-1)}(x)  \big\rangle.
\end{array}
\end{equation}
\end{theorem}

\noindent {\bf Proof :}
Let the ideal on the right hand side  of equation (\ref{eq13}) be denoted as $\mathcal{D}$.
Let $g(x,y,z)$ be an arbitrary element of $\mathcal{C}$. Then there exist polynomials $g_t(x,y) \in \mathbb{F}_q[x,y]/\langle x^s-\alpha, y^\ell-\beta \rangle$ for $t=0,1,\cdots,k-1$ such that $$g(x,y,z)=g_0(x,y)+g_1(x,y)z+\cdots+g_{k-1}(x,y)z^{k-1}.$$
Then using Lemma \ref{lem4},
\begin{equation} \begin{array}{ll} \label{eq14}
g(x,y,z)\zeta_{t}(z)&=g_0(x,y)\zeta_{t}(z)+g_1(x,y)\zeta_{t}(z) z+\cdots+g_{k-1}(x,y)\zeta_{t}(z)z^{k-1} \vspace{2mm}\\
&=g_0(x,y)\zeta_{t}(z)+g_1(x,y)\omega^{1+tr}\zeta_{t}(z)+\cdots+g_{k-1}(x,y)(\omega^{1+tr})^{k-1}
\zeta_{t}(z) \vspace{2mm}\\
&=\zeta_{t}(z)\{g_0(x,y)+g_1(x,y)\omega^{1+tr}+\cdots+g_{k-1}(x,y)(\omega^{1+tr})^{k-1}\} \vspace{2mm}\\
&=\zeta_{t}(z)g(x,y,\omega^{1+tr}).
\end{array} \end{equation}

\noindent Now, $g(x,y,z) \in \mathcal{C}$ implies $g(x,y,z)\zeta_{t}(z) \in \mathcal{C}$ and hence by definition of $I_t$,  $g(x,y,\omega^{1+tr})$ $  \in  I_t$.  As the two-dimensional code $I_t=\langle \eta_0(y) p_0^{(t)}(x), \eta_1(y) p_1^{(t)}(x),\cdots,\eta_{\ell-1}(y) p_{\ell-1}^{(t)}(x) \rangle$ for $t=0,1,\cdots,k-1,$
 there exist some polynomials $ h_j(x,y) \in \mathbb{F}_q[x,y]/\langle x^s-\alpha, y^\ell-\beta \rangle $ such that $$g(x,y,\omega^{1+tr})= \sum_{j=0}^{\ell-1}h_j(x,y) \eta_j(y) p_j^{(t)}(x). $$
From equation (\ref{eq14}), we get that $$g(x,y,z)\zeta_{t}(z)=\zeta_{t}(z) g(x,y,\omega^{1+tr})=\zeta_t(z)\sum_{j=0}^{\ell-1}h_j(x,y) \eta_j(y) p_j^{(t)}(x). $$. \vspace{2mm}

\noindent Since $\displaystyle\sum_{t=0}^{k-1}\zeta_{t}(z)=1$, we get $g(x,y,z)=\displaystyle\sum_{t=0}^{k-1} g(x,y,z)\zeta_{t}(z)=\displaystyle \sum_{t=0}^{k-1} \sum_{j=0}^{\ell-1} \zeta_{t}(z)h_j(x,y) \eta_j(y) p_j^{(t)}(x)$. Therefore,  \begin{equation*}
\begin{array}{ll}
g(x,y,z) \in \big\langle &\zeta_0(z)\eta_0(y)p_0^{(0)}(x),  \cdots , \zeta_0(z)\eta_{\ell-1}(y)p_{\ell-1}^{(0)}(x),\vspace{2mm}\\
&\zeta_1(z)\eta_0(y)p_0^{(1)}(x),  \cdots , \zeta_1(z)\eta_{\ell-1}(y)p_{\ell-1}^{(1)}(x),\\
& \vdots\\
& \zeta_{k-1}(z)\eta_0(y)p_0^{(k-1)}(x),  \cdots , \zeta_{k-1}(z)\eta_{\ell-1}(y)p_{\ell-1}^{(k-1)}(x)  \big\rangle.
\end{array}
\end{equation*}
$i.e.$ $g(x,y,z) \in \mathcal{D}$.
Therefore, $ \mathcal{C} \subseteq \mathcal{D} $
Also, as $\eta_j(y)p_j^{(t)}(x)\in I_t$ for every $j=0,1,\cdots,\ell-1$, we have, by definition $\zeta_t(z)\eta_j(y)p_j^{(t)}(x)\in \mathcal{C}$.  This being true for every $t$, $t= 0,1,\cdots,k-1$, we have $\mathcal{D}= \mathcal{C}.$
$\hfill\Box$

\begin{theorem} \label{th2}
	Let deg $p_j^{(t)}(x)=a_{t,j}$ for $t=0,1,\cdots,k-1$ and $ j=0,1,\cdots,\ell-1 $, then a generator matrix of $\mathcal{C}$ is $$ G=\begin{pmatrix}
	p_0^{(0)}(x) \eta_0(y) \zeta_0(z)\\
	xp_0^{(0)}(x) \eta_0(y) \zeta_0(z)\\
	\vdots\\
	x^{s-a_{0,0}-1}p_0^{(0)}(x) \eta_0(y) \zeta_0(z)\\
	\vdots \vdots\\
	p_{\ell-1}^{(0)}(x) \eta_{\ell-1}(y) \zeta_0(z)\\
	xp_{\ell-1}^{(0)}(x) \eta_{\ell-1}(y) \zeta_0(z)\\
	\vdots\\
	x^{s-a_{0,\ell-1}-1}p_{\ell-1}^{(0)}(x) \eta_{\ell-1}(y) \zeta_0(z)\\
	\vdots\vdots\\
	\vdots\vdots\\
	p_{\ell-1}^{(k-1)}(x) \eta_{\ell-1}(y) \zeta_{k-1}(z)\\
	xp_{\ell-1}^{(k-1)}(x) \eta_{\ell-1}(y) \zeta_{k-1}(z)\\
	\vdots\\
	x^{s-a_{k-1,\ell-1}-1}p_{\ell-1}^{(k-1)}(x) \eta_{\ell-1}(y) \zeta_{k-1}(z)
	\end{pmatrix}.	$$
\end{theorem}

\noindent {\bf Proof :}
It is enough to prove that rows of $G$ are linearly independent. Suppose, if possible, there exist polynomials $m_{t,j}(x)$ in $\mathbb{F}_q[x]/\langle x^s-\alpha \rangle $, with  $\deg m_{t,j}(x) \leq s-a_{t,j}-1$, for ${t=0,1,2, \cdots ,k-1}$ and ${j=0,1,2, \cdots ,\ell-1}$, such that
\begin{equation*}
\begin{array}{ll}
&m_{0,0}(x)\zeta_0(z)\eta_0(y)p_0^{(0)}(x)+  \cdots + m_{0,\ell-1}(x)\zeta_0(z)\eta_{\ell-1}(y)p_{\ell-1}^{(0)}(x)+
\cdots\\
&m_{k-1,0}(x)\zeta_{k-1}(z)\eta_0(y)p_0^{(k-1)}(x)+  \cdots + m_{k-1,\ell-1}\zeta_{k-1}(z)\eta_{\ell-1}(y)p_{\ell-1}^{(k-1)}(x)=0
\end{array}
\end{equation*}

\noindent  in $\mathbb{F}_q[x,y,z]/\langle x^s-\alpha,y^\ell-\beta,z^k-\gamma \rangle.$\\
\noindent Therefore, there exist polynomials $a(x,y,z) , b(x,y,z), c(x,y,z) \in \mathbb{F}_q[x,y,z]$ such that
\begin{equation} \label{eq15}
\sum_{t=0}^{k-1}\sum_{j=0}^{\ell-1} m_{t,j}(x)\zeta_{t}(z)\eta_j(y)p_j^{(t)}(x)=a(x,y,z)(x^s-\alpha) + b(x,y,z)(y^\ell-\beta)+c(x,y,z)(z^k-\gamma).
\end{equation}

 \noindent Substituting $z=\omega^{1+tr} $ and $y=\theta^{1+jr'}$ in equation (\ref{eq15}) and using equations (\ref{eq9}) and (\ref{eq12}), we get
 \begin{equation}\label{eq16}m_{t,j}(x)p_{j}^{(t)}(x)=a(x,\theta^{1+jr'},\omega^{1+tr})(x^s-\alpha).\end{equation}
 \\
 \noindent Since, $\deg( m_{t,j}(x)p_{j}^{(t)}(x)) \leq s-a_{t,j}-1+a_{t,j} < s$ , comparing coefficients of $x^s,x^{s+1},\cdots$ on both sides of (\ref{eq16}), we find that  coefficients  of $a(x,\theta^{1+jr'},\omega^{1+tr})$ are all zero and hence $a(x,\theta^{1+jr'},\omega^{1+tr})=0$ $i.e.$ $ m_{t,j}(x)p_{j}^{(t)}(x)=0$ in $\mathbb{F}_q[x]$. Therefore, $m_{t,j}(x)=0$ for all $t=0,1,\cdots,k-1$ and $j=0,1,\cdots,\ell-1$.  Thus the rows of $G$ form a generator matrix of $\mathcal{C}$. \hfill $\square$

\begin{cor}\normalfont The dimension of  a 3-D $(\alpha,\beta,\gamma )$-constacyclic code  $\mathcal{C}$ of length $s\ell k$ is given by $s\ell k-\sum_{t=0}^{k-1}\sum_{j=0}^{\ell-1}a_{t,j}   $.\end{cor}

\subsection{Generator matrix of dual code}
We assume here that $\alpha, \beta, \gamma \in \{1,-1\}$ so that  the dual code $\mathcal{C}^{\perp}$ of a 3-D $(\alpha,\beta,\gamma )$-constacyclic code $\mathcal{C}$ is also an ideal  in $\mathbb{F}_q[x,y,z]/\langle x^s-\alpha,y^\ell-\beta,z^k-\gamma \rangle.$ \\
\noindent  As $ \dim(\mathcal{C}) + \dim(\mathcal{C}^{\perp})=s \ell k$, therefore $\dim(\mathcal{C}^\perp)=a_{0,0}+\cdots+a_{0,\ell-1}+\cdots+a_{k-1,0}+\cdots+a_{k-1,\ell-1}$. As $p_j^{(t)}(x)$ are divisors of $x^s-\alpha$, there exist polynomials $q_j^{(t)}(x) \in \mathbb{F}_q[x]$ such that $p_j^{(t)}(x)q_j^{(t)}(x)=x^s-\alpha$ for $t=0,1,\cdots,k-1$ and $j=0,1,\cdots,\ell-1$.  The following theorem gives the generators of the dual code $C^\perp$:

\begin{theorem} \label{th3}
Suppose that $\alpha, \beta, \gamma \in \{1,-1\}.$ Let $\mathcal{C}$ be a 3-D $(\alpha, \beta, \gamma)$ - constacyclic code of length $n=s\ell k$  as given in Theorem \ref{th1}. Then the dual code

\begin{equation}\label{eq17}
\begin{array}{ll}
\mathcal{C^\perp} = \big\langle &\zeta_0^*(z)\eta_0^*(y)q_0^{(0)*}(x),  \cdots , \zeta_0^*(z)\eta_{\ell-1}^*(y)q_{\ell-1}^{(0)*}(x),\\
&\zeta_1^*(z)\eta_0^*(y)q_0^{(1)*}(x),  \cdots , \zeta_1^*(z)\eta_{\ell-1}^*(y)q_{\ell-1}^{(1)*}(x),\\
& \vdots\\
& \zeta_{k-1}^*(z)\eta_0^*(y)q_0^{(k-1)*}(x),  \cdots , \zeta_{k-1}^*(z)\eta_{\ell-1}^*(y)q_{\ell-1}^{(k-1)*}(x)  \big\rangle.
\end{array}
\end{equation}
and a generator matrix of $\mathcal{C}^\perp$ is given by,
$$ H=\begin{pmatrix}
\zeta_0^*(z)\eta_0^*(y)q_0^{(0)*}(x)\\
x\zeta_0^*(z)\eta_0^*(y)q_0^{(0)*}(x)\\
\vdots\\
x^{a_{0,0}-1}\zeta_0^*(z)\eta_0^*(y)q_0^{(0)*}(x)\\
\vdots \vdots\\
\zeta_0^*(z)\eta_{\ell-1}^*(y)q_{\ell-1}^{(0)*}(x)\\
x\zeta_0^*(z)\eta_{\ell-1}^*(y)q_{\ell-1}^{(0)*}(x)\\
\vdots\\
x^{a_{0,\ell-1}-1}\zeta_0^*(z)\eta_{\ell-1}^*(y)q_{\ell-1}^{(0)*}(x)\\
\vdots\vdots\\
\vdots\vdots\\
\zeta_{k-1}^*(z)\eta_{\ell-1}^*(y)q_{\ell-1}^{(k-1)*}(x)\\
x\zeta_{k-1}^*(z)\eta_{\ell-1}^*(y)q_{\ell-1}^{(k-1)*}(x)\\
\vdots\\
x^{a_{k-1,\ell-1}-1}\zeta_{k-1}^*(z)\eta_{\ell-1}^*(y)q_{\ell-1}^{(k-1)*}(x)
\end{pmatrix},	$$
where $\zeta_{t}^*(z), \eta_{j}^*(y)$ and $q_{j}^{(t)*}(x)$ denote the reciprocal polynomials of $\zeta_{t}(z), \eta_{j}(y)$ and $q_{j}^{(t)}(x)$ respectively.
\end{theorem}

\noindent {\bf Proof:} Let the code on the right hand side of equation (\ref{eq17}) be denoted by $D$. To prove that $D \subset \mathcal{C}^\perp$ it is enough to prove, by Proposition 4, that
$\zeta_t^*(z)\eta_j^*(y)q_j^{(t)*}(x)\in {\rm ann}(\mathcal{C})^\ast$ i.e, $\zeta_t(z)\eta_j(y)q_j^{(t)}(x)\in {\rm ann}(\mathcal{C})$ for each $t,j;~0\le t \le k-1,~ 0\le j \le \ell-1$.  As $\mathcal{C}=\langle \zeta_u(z)\eta_v(y)p_v^{(u)}(x), ~0\leq u \leq k-1, ~0\leq v\leq \ell-1\rangle$, it is enough to prove that  
 \begin{equation}\label{eq18}\zeta_t(z)\eta_j(y)q_j^{(t)}(x)\zeta_u(z)\eta_v(y)p_v^{(u)}(x)=0 \end{equation} for all $u,v;~0\le u \le k-1,~ 0\le v \le \ell-1$.
  Now if $u \neq t$, we have $\zeta_u(z)\zeta_t(z)=0$; when $v \neq j$, we have $\eta_v(y)\eta_j(y)=0$ and when $u=t$, $v=j$ we have $p_j^{(t)}(x) q_j^{(t)}(x)=x^s-\alpha=0$ in the ring $\mathcal{R}$. Hence the expression (\ref{eq18}) holds i.e. $D \subset \mathcal{C}^\perp$. \vspace{2mm}

To prove that   $\dim(D)=a_{0,0}+\cdots+a_{0,\ell-1}+\cdots+a_{k-1,0}+\cdots+a_{k-1,\ell-1}=\dim (\mathcal{C}^\perp)$, we need to show that  the rows of $H$  are linearly independent.
 Suppose, if possible, there exist polynomials $m_{0,0}(x),\cdots, m_{0,\ell-1}(x), \cdots ,m_{k-1,\ell-1}(x)$ in $\mathbb{F}_q[x]/\langle x^s-\alpha \rangle $, with $\deg m_{t,j}(x) \leq a_{t,j}-1$ for ${t=0,1,2, \cdots ,k-1}$ and ${j=0,1,2, \cdots ,\ell-1}$ such that $$m_{0,0}(x)q_0^{(0)*}(x) \eta_0^*(y) \zeta_0^*(z) +  \cdots + m_{k-1,\ell-1}(x)q_{\ell-1}^{(k-1)*}(x) \eta_{\ell-1}^*(y) \zeta_{k-1}^*(z)=0$$ in  $\mathbb{F}_q[x,y,z]/\langle x^s-\alpha,y^\ell-\beta,z^k-\gamma \rangle.$\\
 \noindent Therefore, there exist polynomials $a(x,y,z) , b(x,y,z), c(x,y,z) \in \mathbb{F}_q[x,y,z]$ such that
\begin{equation} \label{eq19}
\sum_{t=0}^{k-1}\sum_{j=0}^{\ell-1}m_{t,j}(x)q_j^{(t)*}(x) \eta_j^*(y)\zeta_t^*(z)=a(x,y,z)(x^s-\alpha) + b(x,y,z)(y^\ell-\beta) +c(x,y,z)(z^k-\gamma).
\end{equation}

\noindent Let first $r=r'=1$. Then by Lemma \ref{lem5}, $\zeta_t^\ast(z) = c_t~\zeta_{k-2-t}(z)$ and $\eta_j^\ast(y) = b_j~\eta_{\ell-2-j}(y)$ for some constant $c_t$ and $b_j \in \mathbb{F}_q^\ast$. Also by equation (\ref{eq9}), we have $\zeta_{k-2-t}(\omega^{k-1-t})=1$, and $\zeta_{k-2-t'}(\omega^{k-1-t})=0$ for $t\neq t'$. Similarly by (\ref{eq12}), we have $\eta_{\ell-2-j}(\theta^{\ell-1-j})=1$, and $\eta_{\ell-2-j'}(\theta^{\ell-1-j})=0$ for $j\neq j'$.\vspace{2mm}

 \noindent Substituting $ y=\theta^{\ell-1-j} $, $z=\omega^{k-1-t} $ in equation (\ref{eq19}) , we get
 $$m_{t,j}(x)  c_{t} b_{j} q_{j}^{(t)*}(x)=a(x,\theta^{\ell-1-j},\omega^{k-1-t})(x^s-\alpha).$$
  Since $\deg\big( m_{t,j}(x)q_{j}^{(t)*}(x)\big) \leq a_{t,j}-1+s-a_{t,j} < s$, comparing coefficients of $x^s,x^{s+1},\cdots$ on both sides we find that  coefficients  of $a(x,\theta^{\ell-1-j},\omega^{k-1-t})$ are all zero and hence $a(x,\theta^{\ell-1-j},\omega^{k-1-t})=0$ in $\mathbb{F}_q[x]$. Therefore, $m_{t,j}(x)=0$ for all $j=0,1,\cdots,\ell-1$ and $t=0,1,\cdots,k-1$. \vspace{2mm}

\noindent Let now $r=r'=2$. Then by Lemma \ref{lem5}, $\zeta_t^\ast(z) = c_t\zeta_{k-1-t}(z)$ and $\eta_j^\ast(y) = b_j~\eta_{\ell-1-j}(y)$. Also by equation (\ref{eq9}), we have $\zeta_{k-1-t}(\omega^{1+2(k-1-t)})=1$, and $\zeta_{k-1-t'}(\omega^{1+2(k-1-t)})=0$ for $t\neq t'$. Similarly, by (\ref{eq12}), we have $\eta_{\ell-1-j}(\theta^{1+2(\ell-1-j)})=1$, and $\eta_{\ell-1-j'}(\theta^{1+2(\ell-1-j)})=0$ for $j\neq j'$.
 Substituting $ y=\theta^{1+2(\ell-1-j)} $, $z=\omega^{1+2(k-1-t)} $ in equation (\ref{eq16}) , we get
 $$m_{t,j}(x)  c_{t} b_{j} q_{j}^{(t)*}(x)=a(x,\theta^{1+2(\ell-1-j)},\omega^{1+2(k-1-t)})(x^s-\alpha).$$
  Working as above we find that $m_{t,j}(x)=0$ for all $j=0,1,\cdots,\ell-1$ and $t=0,1,\cdots,k-1$.

 \noindent When $r=2$ and $r'=1$.  By Lemma \ref{lem5}, $\zeta_t^\ast(z) = c_t\zeta_{k-1-t}(z)$ and $\eta_j^\ast(y) = b_j~\eta_{\ell-2-j}(y)$. Also by equations (\ref{eq9}) and (\ref{eq12}), we have $\zeta_{k-1-t}(\omega^{1+2(k-1-t)})=1$, $\zeta_{k-1-t'}(\omega^{1+2(k-1-t)})=0$ for $t\neq t'$;  $\eta_{\ell-2-j}(\theta^{\ell-1-j})=1$, and $\eta_{\ell-2-j'}(\theta^{\ell-1-j})=0$ for $j\neq j'$.
 We substitute $ y=\theta^{\ell-1-j} $, $z=\omega^{1+2(k-1-t)} $ in equation (\ref{eq16}) and
  working as above  find that $m_{t,j}(x)=0$ for all $j=0,1,\cdots,\ell-1$ and $t=0,1,\cdots,k-1$.

  \noindent The case $r=1$ and $r'=2$ is similar.

%
%

\noindent Therefore $D=\mathcal{C}^\perp$ and the rows of $H$ form a generator matrix of $\mathcal{C}^\perp$.
 \hfill $\square$\vspace{2mm}

\noindent{\bf Remark} If $\deg(p_j^{(t)}(x))=a_{t,j}=0$ for some $t,j;~0\le t \le k-1 , 0\le j \le \ell-1$, i.e. $p_j^{(t)}(x)=\lambda$, a constant, then the polynomials $$\zeta_{t}^*(z)\eta_{j}^*(y)q_{j}^{(t)*}(x),
x\zeta_{t}^*(z)\eta_{j}^*(y)q_{j}^{(t)*}(x),\cdots,
x^{a_{t,j}-1}\zeta_{t}^*(z)\eta_{j}^*(y)q_{j}^{(t)*}(x)$$ do not contribute any rows in $H$.

 \subsection{Self-dual codes}
\begin{theorem}\label{th4} Suppose that $\alpha, \beta, \gamma \in \{1,-1\}$. Then a three-dimensional $(\alpha, \beta, \gamma)$ - constacyclic code $\mathcal{C}$ of length $n=s\ell k$  is self-dual if and only if
\begin{enumerate}[$\rm(i)$] \item $s\ell k = 2(a_{00}+\cdots+a_{0,\ell-1}+\cdots+a_{k-1,0}+\cdots+a_{k-1,\ell-1})$ \item   for every $t,j; 0\le t \le k-1, 0\le j \le \ell-1$,
	\begin{equation}\label{eq20}\begin{array}{lll}
		q_j^{(t)*}(x) = m_{tj}(x) p_{\ell-2-j}^{(k-2-t)}(x),~ p_j^{(t)}(x) = m_{tj}'(x) q_{\ell-2-j}^{(k-2-t)*}(x) &{\rm ~if~}& \beta =1, \gamma =1\end{array}
	\vspace{-1cm}\end{equation}
		
\begin{equation*}\begin{array}{lll}		q_j^{(t)*}(x) = m_{tj}(x) p_{\ell-1-j}^{(k-1-t)}(x),~ p_j^{(t)}(x) = m_{tj}'(x) q_{\ell-1-j}^{(k-1-t)*}(x) &{\rm ~if~}& \beta =-1, \gamma =-1\vspace{3mm}\\
		
		q_j^{(t)*}(x) = m_{tj}(x) p_{\ell-2-j}^{(k-1-t)}(x),~ p_j^{(t)}(x) = m_{tj}'(x) q_{\ell-2-j}^{(k-1-t)*}(x) &{\rm ~if~}& \beta =1, \gamma =-1\vspace{3mm}\\
		
		q_j^{(t)*}(x) = m_{tj}(x) p_{\ell-1-j}^{(k-2-t)}(x),~ p_j^{(t)}(x) = m_{tj}'(x) q_{\ell-1-j}^{(k-2-t)*}(x) &{\rm ~if~}& \beta =-1, \gamma =1
	\end{array}
	\end{equation*}
for some non-zero polynomials  $m_{tj}(x), m_{tj}'(x)$ in $ \mathbb{F}_q[x]/\langle x^s-\alpha \rangle $. When $\gamma=1$ and $t=k-1$, the superscript $k-2-t$ be replaced by $k-1$. Similarly when $\beta =1$ and $j=\ell-1$, the subscript $\ell-2-j$ be replaced by $\ell-1$.\end{enumerate}
\end{theorem}

\noindent {\bf Proof:}  It is clear that if $\mathcal{C}$ is self-dual, then $s\ell k=2(a_{00}+\cdots+a_{0,\ell-1}+\cdots+a_{k-1,0}+\cdots+a_{k-1,\ell-1})$. By Theorems \ref{th1} and \ref{th3},
$$\mathcal{C} = \big\langle p_j^{(t)}(x)\eta_j(y)\zeta_t(z): 0\leq t \leq k-1, 0\leq j \leq \ell-1\big\rangle$$
 and
$$\mathcal{C^\perp} = \big\langle  q_j^{(t)*}(x)\eta_j^*(y)\zeta_t^*(z): 0\leq t \leq k-1, 0\leq j \leq \ell-1 \big\rangle.$$
\noindent Therefore $\mathcal{C}^\perp \subset \mathcal{C}$ if  each  $q_j^{(t)*}(x)\eta_j^*(y)\zeta_t^*(z)$ is a linear combination of rows of generator matrix $G$ of $\mathcal{C}$ as given in Theorem \ref{th2}. This is so if and only if there exist polynomials $h_{uv}(x)$ in $ \mathbb{F}_q[x]/\langle x^s-\alpha \rangle $ of $\deg h_{uv}\leq s-a_{u,v}-1$ such that
\begin{equation}\label{eq21} q_j^{(t)*}(x)\eta_j^*(y)\zeta_t^*(z)=\sum_{u=0}^{k-1}\sum_{v=0}^{\ell-1}h_{uv}(x)p_v^{(u)}(x)\eta_v(y)\zeta_u(z).\end{equation}

\noindent Again $ \mathcal{C} \subset \mathcal{C^\perp} $, if each $p_j^{(t)}(x) \eta_j(y) \zeta_t(z) $ is a linear combination of rows of  generator matrix $H$ of $\mathcal{C^\perp}$ as given in Theorem \ref{th3}. This is so if and only if there exists polynomials $h'_{uv}(x)$ in $\mathbb{F}_q[x]/ \langle x^s-\alpha \rangle$ of $\deg h'_{uv}\leq a_{u,v}-1$ such that
\begin{equation}\label{eq22} p_j^{(t)}(x)\eta_j(y)\zeta_t(z)=\sum_{u=0}^{k-1}\sum_{v=0}^{\ell-1}h'_{uv}(x)q_v^{(u)*}(x)\eta_v^*(y)\zeta_u^*(z).\end{equation}

\noindent When $\beta =1=\gamma$, we have $r=r'=1$ and $\zeta_t^*(z)=b_t \zeta_{k-2-t}(z), \eta_j^*(y)=c_j \eta_{\ell-2-j}(y)$ by lemmas \ref{lem5} and \ref{lem6}. When  $t=k-1$, $\zeta_{k-2-t}(z)=\zeta_{k-1}(z)$ and  when $j=\ell-1$,  $\eta_{\ell-2-j}(y)=\eta_{\ell-1}(y)$. Multiplying both sides of (\ref{eq21}) by $\zeta_{k-2-t}(z)\eta_{\ell-2-y}(y)$ we get
$$q_j^{(t)*}(x) c_j \eta_{\ell-2-j}^2(y) b_t \zeta_{k-2-t}^2(z)= \sum_{u=0}^{k-1}\sum_{v=0}^{\ell-1}h_{uv}(x)p_v^{(u)}(x)\eta_v(y)\eta_{\ell-2-j}(y)\zeta_u(z)\zeta_{k-2-t}(z). $$
\noindent As $\zeta_u(z), \eta_v(y)$ are primitive central idempotents, $\mathcal{C^\perp} \subset \mathcal{C}$ if and only if
$$ q_j^{(t)*}(x) c_j \eta_{\ell-2-j}(y) b_t \zeta_{k-2-t}(z)=h_{k-2-t,\ell-2-j}(x) p_{\ell-2-j}^{(k-2-t)}(x)\eta_{\ell-2-j}(y)  \zeta_{k-2-t}(z)$$
\noindent i.e. if and only if
$$ q_j^{(t)*}(x) =m_{tj}(x) p_{\ell-2-j}^{(k-2-t)}(x)$$
\noindent for some polynomial $m_{tj}(x)$.\\
\noindent Further, (\ref{eq22}) can be rewritten as
\begin{equation}\label{eq23}\begin{array}{ll}
p_j^{(t)}(x)\eta_j(y)\zeta_t(z) &=\displaystyle \sum_{u=0}^{k-1}\sum_{v=0}^{\ell-1}b_u c_v h'_{uv}(x) q_v^{(u)*}(x) \eta_{\ell-2-v}(y)\zeta_{k-2-u}(z)\vspace{3mm}\\
&=\displaystyle\sum_{u=-1}^{k-2}\displaystyle\sum_{v=-1}^{\ell-2}b_{k-2-u} c_{\ell-2-v} h'_{k-2-u, \ell-2-v}(x) q_{\ell-2-v}^{(k-2-u)*}(x) \eta_v(y)\zeta_u(z)\\
\end{array}
\end{equation}
with the understanding that $\zeta_{-1}(z)=\xi_0(z)=\zeta_{k-1}(z)$ from equation (\ref{eq8}) and similarly $\eta_{-1}(y)=\eta_{\ell-1}(y)$. \vspace{2mm}

\noindent Multiplying both sides of (\ref{eq23}) by $\eta_j(y) \zeta_t(z)$, $\mathcal{C} \subset \mathcal{C^\perp}$ if and only if
$$ p_j^{(t)}(x)\eta_j(y)\zeta_t(z) = b_{k-2-t} c_{\ell-2-j} h'_{k-2-t, \ell-2-j}(x) q_{\ell-2-j}^{(k-2-t)*}(x) \eta_j(y)\zeta_t(z)$$
\noindent i.e. if and only if
$$ p_j^{(t)}(x) = m_{tj}'(x) q_{\ell-2-j}^{(k-2-t)*}(x)$$
\noindent for some polynomial $m_{tj}'(x)$.\\

When $\beta=-1=\gamma$, we have $r=r'=2$, $\zeta_t^*(z)=b_t \zeta_{k-1-t}(z), \eta_j^*(y)=c_j \eta_{\ell-1-j}(y)$; when $\beta=1, \gamma=-1$, we have  $r=2, r'=1$, $\zeta_t^*(z)=b_t \zeta_{k-1-t}(z), \eta_j^*(y)=c_j \eta_{\ell-2-j}(y)$; and when $\beta=-1, \gamma=1$, we have  $r=1, r'=2$, $\zeta_t^*(z)=b_t \zeta_{k-2-t}(z), \eta_j^*(y)=c_j \eta_{\ell-1-j}(y)$ and the proof is similar.

\begin{theorem}\label{th5} If $\beta=1=\gamma$ and $\alpha=\pm 1$ then  a three-dimensional $(\alpha,1,1 )$-constacyclic code $\mathcal{C}$ of length $n=s\ell k$  can not be self-dual   if $\gcd(s,q)= 1$, assuming that  $s$ is odd if $\alpha=-1$. \end{theorem}

\noindent {\bf Proof : }   Suppose a three-dimensional $(\alpha,1,1 )$-constacyclic code is self-dual. On taking $t=k-1, ~j=\ell-1$ in equation (\ref{eq20}) we get
\begin{equation}\label{eq24}
q_{\ell -1}^{(k-1)*}(x) = m_{k-1,\ell -1}(x) p_{\ell-1}^{(k-1)}(x),\end{equation}
and
\begin{equation}\label{eq25}~ p_{\ell-1}^{(k-1)}(x) = m_{k-1,\ell-1}'(x) q_{\ell-1}^{(k-1)*}(x).\end{equation}

\noindent We have $x^s-\alpha=p_{\ell-1}^{(k-1)}(x)q_{\ell-1}^{(k-1)}(x)$. If $\gcd(s,q)= 1$,  $x-\alpha$  divides exactly one of $p_{\ell-1}^{(k-1)}(x)$ and $q_{\ell-1}^{(k-1)}(x)$ and not both (If $\alpha =-1$, we assume that $s$ is odd). The reciprocal of $x-\alpha$ is $\pm(x-\alpha)$.\vspace{2mm}

\noindent If $x-\alpha | p_{\ell-1}^{(k-1)}(x)$, then from equation (\ref{eq24}) we have $x-\alpha | q_{\ell -1}^{(k-1)*}(x)$. This implies
$(x-\alpha)^* | (q_{\ell -1}^{(k-1)*}(x))^*$ i.e. $x-\alpha | q_{\ell -1}^{(k-1)}(x)$ as $(f^*)^*=f$. This is not possible, when $\gcd(s,q)= 1$.  \vspace{2mm}

\noindent If $x-\alpha | q_{\ell-1}^{(k-1)}(x)$,  we have $(x-\alpha)^* | q_{\ell-1}^{(k-1)*}(x)$ i.e. $(x-\alpha) | q_{\ell-1}^{(k-1)*}(x)$. This implies, from equation (\ref{eq25}),
$x-\alpha | p_{\ell-1}^{(k-1)}(x)$, again not possible.

\subsection{Examples}
Minimum distances of  codes in the following examples have been calculated by software MAGMA.
\begin{eg}
	\normalfont	Let $q=5$,  $\alpha =1$ , $\beta =-1$ , $\gamma=-1$ , $s=2=\ell=k$. One finds that $\omega=2 $ is a $4^{th}$ root of unity in $\mathbb{F}_{5}^\ast$ such that $\omega ^{2} = -1$. Therefore
	\begin{equation*}
	\begin{array}{ll}
	z^2+1 &=(z-2)(z-2^3) \vspace{2mm}\\
	&=(z-2)(z-3).
	\end{array}
	\end{equation*}
	
	\noindent Thus,
	\begin{equation*}
	\begin{array}{ll}
	\zeta_{0}(z) &=-z+3, \vspace{2mm}\\
	\zeta_{1}(z) &=z+3, \vspace{2mm}\\
	\end{array}
	\end{equation*}
	Also $\theta=2 $ is a $4^{th}$ root of unity in $\mathbb{F}_{5}^\ast$ such that $\theta ^{2} = -1$. Therefore,
	\begin{equation*}
	\begin{array}{ll}
	\eta_{0}(y) &=-y+3, \vspace{2mm}\\
	\eta_{1}(y) &=y+3, \vspace{2mm}\\
	\end{array}
	\end{equation*}
 \noindent We have $x^2-1=(x-1)(x+1)$. Suppose, $p_0^{(0)}(x)=p_0^{(1)}(x)=x-1$, and $p_1^{(0)}(x)=p_1^{(1)}(x)=x+1$, then by Theorems \ref{th2} and \ref{th3},  generator matrices  of two dimensional $ \left( 1,-1,-1 \right) $-constacyclic code $\mathcal{C}$ and $\mathcal{C}^\perp$ are given by
$$
G=\begin{pmatrix}
	(x-1)(-y+3)(-z+3)\\
(x+1)(y+3)(-z+3)\\
(x-1)(-y+3)(z+3)\\
(x+1)(y+3)(z+3)
\end{pmatrix},H=\begin{pmatrix}
(x+1)(-1+3y)(-1+3z)\\
(-x+1)(1+3y)(-1+3z)\\
(x+1)(-1+3y)(1+3z)\\
(-x+1)(1+3y)(1+3z)
\end{pmatrix}.
$$
\noindent respectively. The corresponding code $\mathcal{C}_3$ has a generator matrix
$$G_3=\begin{pmatrix}
1&-1&-2&2&-2&2&-1&1\\
-1&-1&-2&-2&2&2&-1&-1\\
1&-1&-2&2&2&-2&1&-1\\
-1&-1&-2&-2&-2&-2&1&1
\end{pmatrix}.
$$

\noindent Then $\mathcal{C}_3$ is a  $[8,4,2]$ self-dual code over $\mathbb{F}_{5}[x]$. It is a (-1)-quasi-twisted code of index 2. \vspace{2mm}

\end{eg}
\begin{eg}
	\normalfont	Let $q=7$,  $\alpha =1$ , $\beta =1$ , $\gamma=-1$ , $s=2=\ell$ and $k=3$. One finds that $\omega=3 $ is a $6^{th}$ root of unity in $\mathbb{F}_{7}^\ast$ such that $\omega ^{3} = -1$. Therefore
	\begin{equation*}
	\begin{array}{ll}
	z^3+1 &=(z-3)(z-3^3(z-3^5)) \vspace{2mm}\\
	&=(z-3)(z+1)(z+2).
	\end{array}
	\end{equation*}
	
	\noindent Thus,
	\begin{equation*}
	\begin{array}{ll}
	\zeta_{0}(z) &=-z^2-3z-2, \vspace{2mm}\\
	\zeta_{1}(z) &=-2z^2+2z-2, \vspace{2mm}\\
	\zeta_{2}(z) &=3z^2+z-2, \vspace{2mm}\\
	\end{array}
	\end{equation*}
	Also $y^2-1=(y-1)(y+1)$, hence
	\begin{equation*}
	\begin{array}{ll}
	\eta_{0}(y) &=-3y-3, \vspace{2mm}\\
	\eta_{1}(y) &=3y-3, \vspace{2mm}\\
	\end{array}
	\end{equation*}

	\noindent We have $x^2-1=(x-1)(x+1)$. Suppose, $p_0^{(0)}(x)=p_0^{(1)}(x)=p_0^{(2)}(x)=x-1$, and $p_1^{(0)}(x)=p_1^{(1)}(x)=p_1^{(2)}(x)=x+1$, then by Theorems \ref{th2} and \ref{th3},  generator matrices  of two dimensional $ \left( 1,1,-1 \right) $-constacyclic code $\mathcal{C}$ and $\mathcal{C}^\perp$ are given by
	$$
	G=\begin{pmatrix}
	(x-1)(-3y-3)(-z^2-3z-2)\\
	(x+1)(3y-3)(-z^2-3z-2)\\
	(x-1)(-3y-3)(-2z^2+2z-2)\\
	(x+1)(3y-3)(-2z^2+2z-2)\\
	(x-1)(-3y-3)(3z^2+z-2)\\
	(x+1)(3y-3)(3z^2+z-2)
	\end{pmatrix},H=\begin{pmatrix}
	(x+1)(-3y-3)(-2z^2-3z-1)\\
	(-x+1)(-3y+3)(-2z^2-3z-1)\\
	(x+1)(-3y-3)(-2z^2+2z-2)\\
	(-x+1)(-3y+3)(-2z^2+2z-2)\\
	(x+1)(-3y-3)(-2z^2+z+3)\\
	(-x+1)(-3y+3)(-2z^2+z+3)
	\end{pmatrix}.
	$$
	\noindent respectively. By Theorem \ref{th4}, the code $\mathcal{C}$ can not be  self-dual as $p_j^{(t)}=p_0^{(0)}=x-1$ but $q_{\ell-2-j}^{(k-2-t)*}=q_0^{(1)*}=x+1.$ The corresponding code $\mathcal{C}_3$ is a  $[12,6,2]$  code over $\mathbb{F}_{7}[x]$. It is a (-1)-quasi-twisted code of index 3. \vspace{2mm}
	
\end{eg}

\begin{eg}
	\normalfont	Let $q=7$,  $\alpha =-1$ , $\beta =2$ , $\gamma=-1$ , $s=3$ , $\ell=2$ and $k =3$. As computed in Example 2, we have
	
	\begin{equation*}
	\begin{array}{lll}
	\zeta_{0}(z) &=-z^2-3z-2&=6z^2+4z+5, \vspace{2mm}\\
	\zeta_{1}(z) &=-2z^2+2z-2&=5z^2-5z+5, \vspace{2mm}\\
	\zeta_{2}(z) &=3z^2+z-2&=3z^2+z+5, \vspace{2mm}\\
	\end{array}
	\end{equation*}
	Also $\theta=3 $ is a $6^{th}$ root of unity in $\mathbb{F}_{7}^\ast$ such that $\theta ^{2} = 2$. Therefore
	\begin{equation*}
	\begin{array}{ll}
	y^2-2 &=(y-3)(y-3^4) \vspace{2mm}\\
	&=(y-3)(y-4).
	\end{array}
	\end{equation*}
	
	\noindent And hence,
	\begin{equation*}
	\begin{array}{ll}
	\eta_{0}(y) &=6y+4, \vspace{2mm}\\
	\eta_{1}(y) &=y+4, \vspace{2mm}\\
	\end{array}
	\end{equation*}

	\noindent We have $x^3+1=(x+1)(x^2-x+1)$. Suppose, $p_0^{(0)}(x)=p_0^{(1)}(x)=x^2-x+1$ , $p_1^{(0)}(x)=p_0^{(2)}(x)=x+1$ and $p_1^{(1)}(x)=p_1^{(2)}(x)=1$, then by Theorem \ref{th2},   generator matrix  of three dimensional $ \left( -1,2,-1 \right) $-constacyclic code $\mathcal{C}$ is given by
	$$
	G=\begin{pmatrix}
	(x^2-x+1)(6y+4)(6z^2+4z+5)\\
	(x+1)(y+4)(6z^2+4z+5)\\
	x(x+1)(y+4)(6z^2+4z+5)\\
	(x^2-x+1)(6y+4)(5z^2-5z+5)\\
	1(y+4)(5z^2-5z+5)\\
	x(y+4)(5z^2-5z+5)\\
	x^2(y+4)(5z^2-5z+5)\\
	(x+1)(6y+4)(3z^2+z+5)\\
	x(x+1)(6y+4)(3z^2+z+5)\\
	1(y+4)(3z^2+z+5)\\
	x(y+4)(3z^2+z+5)\\
	x^2(y+4)(3z^2+z+5)
	\end{pmatrix}.
	$$
	\noindent The corresponding code $\mathcal{C}_3$  is a  $[18,12,4]$  code over $\mathbb{F}_{7}[x]$. It is a (-1)-quasi-twisted code of index 3. \vspace{2mm}
	
\end{eg}
\section{Conclusion}
In this paper we  characterize the algebraic structure of three-dimensional $(\alpha,\beta,\gamma)$- constacyclic codes of arbitrary length $s\ell k$ and their duals over a finite field  $\mathbb{F}_q$, where $\alpha,\beta,\gamma$ are non zero elements of $\mathbb{F}_q$. We give necessary and sufficient conditions for a three-dimensional constacyclic code to be self-dual. The same technique can be applied to characterize the algebraic structure of multi-dimensional  constacyclic codes  and their duals over a finite field  $\mathbb{F}_q$.

\end{document}